\documentclass[twocolumn,aps,prl,showpacs,floatfix,superscriptaddress]{revtex4}
\usepackage{epsfig}
\usepackage{bm}
\usepackage{color}

\usepackage[latin1]{inputenc}
\newcommand{\be}{\begin{equation}}
\newcommand{\ee}{\end{equation}}
\newcommand{\bea}{\begin{eqnarray}}
\newcommand{\eea}{\end{eqnarray}}

\usepackage{amsmath}
\usepackage{amsfonts}
\usepackage{amssymb}
\usepackage{bm}
\usepackage{color}
\usepackage{graphicx}

\newcommand{\la}{\langle}
\newcommand{\ra}{\rangle}

\begin{document}

\title{Stochastic Resonance: from climate to biology}

\author{Roberto Benzi} \affiliation{Dipartimento di Fisica and INFN,
  Universit\`a di Roma ``Tor Vergata'', Via della Ricerca Scientifica
  1, 00133 Roma, Italy.}

\date{\today}
\begin{abstract}
In this paper I will review some basic aspects of the mechanism of stochastic resonance.
Stochastic resonance was first introduced as a possible mechanism to explain long term climatic
variation. Since then, there have been many applications of stochastic resonance in physical and
biological systems. I will show that in complex system, stochastic resonance can substantially change
as a function of the ``system complexity''. Also, I will briefly mention how to apply stochastic resonance
for the case of Brownian motors.
\end{abstract}

\maketitle
\section{1. Introduction}

The first numerical simulation providing strong evidence of Stochastic Resonance was performed
by Angelo Vulpiani and myself in a rather exciting night on February 1980. Together with Alfonso Sutera
and Giorgio Parisi, we were trying, at that time, to understand whether a relatively small periodic forcing can be 
amplified by internal non linear stochastic dynamics, leading to a possible understanding of the
100 Kyear cycle observed in climate records. The first version of our paper was not accepted for
publication in Journal of Geophyisical Research and in Journal of Atmospheric Science.
Eventually the paper  was published
in Tellus in 1982 \cite{TELLUS} together with a similar paper by Nicolis and Nicolis \cite{NICOLIS} 
who, independently, were proposing
the same mechanism for climatic change. Short after that february night in 1980, we were able to provide
a quite general understanding how the mechanism  works and how to generalize it for chaotic systems.
Both theory and generalization appeared as a letter in J. Phys A in 1981 \cite{JPA}, i.e. one year before
the Tellus paper!

We were, and still we are, convinced that Stochastic Resonance  is a rather new and conceptually important
phenomenon in science. For almost a decade, only few colleagues were shearing with us the same feeling. 
Eventually, after the paper of Moss et.al. \cite{MOSS} and some new theoretical work \cite{MCNAMARA}
stochastic resonance became a fashionable and
interesting research topic in many different scientific areas, see \cite{RMP} for a review.

Resonance is a kind of magic word in physics and there is no surprise that a new resonance mechanism can
excite the scientific community. In our case, the name stochastic resonance was introduce because
of a short discussion with J.Imbrie on march 1980. Just few days after our first numerical simulation, 
I participated to a climate meeting in Erice where I gave a short talk on our results. J.Imbrie, one
of the most famous scientist working on paleoclimate, asked whether what we found was somehow
similar to a resonance and my answer was: ``not exactly! It is a kind of stochastic resonance!''. This
is how the name was introduced in literature. In some sense, we can think of a ``resonance'' as follows.
In the standard resonance mechanism, think for instance of a damped harmonic oscillator, the amplification
of the external forcing can be related, mathematically, to a singularity in the complex plane of the
green function of the problem. The real part of the singularity is, of course, the resonance frequency. In chaotic or
stochastic systems, there are singularities in the complex plane but on the imaginary axis. The mechanism
of stochastic resonance provides a way to shift the singularity on the complex plane with a non zero real
part. This rather cumbersome view of stochastic resonance may be not entirely clear but it tries to justify why the
word ``resonance'' can still be used beside any historical reason.

In this paper, I will review the main idea on stochastic resonance with an emphasis on climate research, where it
was originally proposed. I will also show how stochastic resonance is still an interesting subject to work on and in
particular I will discuss some new findings in simple and complex systems. Finally, I will mention
how mechanisms related to stochastic resonance are of interest in biology and in the wide research field of brownian motors.

\section{2. The mechanism of Stochastic Resonance in climate theory}

Climate is one of the most complex system in nature and it is a major scientific challenge to understand
the basic mechanisms leading to climate changes. In the past, climate shows quite remarkable
changes over all times scales. In particular, one of the most striking aspects of past climate changes
is the so called 100ky cycles observed in paleoclimatic records. The cycle itself is not at all periodic 
showing sudden warming phases followed by gentle decreases of temperature. This saw-tooth behaviour is in phase 
with the so called Milankovitch cycle, i.e. to the change of global radiation of the Sun due to astronomical 
change in Earth orbit. This is the only global (i.e. averaged) change experienced in the  incoming radiation. 
Is that possible to relate the astronomical forcing to the observed climate change? This is the basic question
which puzzles scientific research since many decades. A simple computation shows that it is not easy to
guess the correct answer. 

Let us consider the simplest possible climate model. We consider the averaged Earth temperature $T$ as the
basic variable we want to describe, the precise meaning of  average is of marginal interest for the
time being. We know how much radiation is incoming, let us say $R_{in}$. The out coming radiation $R_{out}$
depends on the earth's temperature $T$ (infrared emission) and reflection from the surface, which can
be written as $\alpha R_{in}$. Assuming $C_E$ the earth's thermal inertia, we can write the energy
balance model as:
\be
C_E \frac{dT}{dt} = R_{in}-R_{out}
\label{climate}
\ee
where $R_{out}= \alpha R_{in} + E_I$, $E_I$ being the infrared emission.
\begin{figure}
\centering
\epsfig{width=.40\textwidth,file=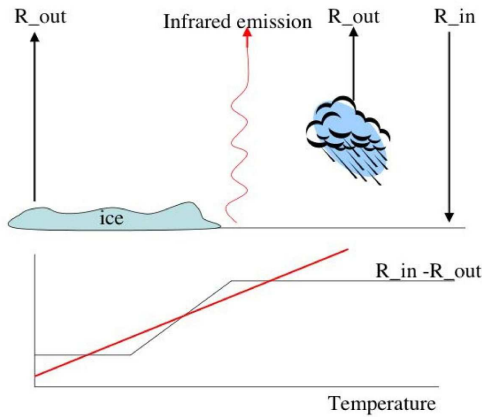}
\caption{The energy balance climate model. In the upper layer we show the basic variables of the model: the incoming 
radiation $R_{in}$ and the out coming radiation $R_{out}$ and the infrared emission. In the lower panel we show the
albedo as a function of temperature $R_{in}-R_{out}= (1-\alpha(T))R_{in}$. The thick line represents the infrared emission.}
\label{fig1}
\end{figure}

What we might call Earth'climate is just the stationary solution of equation (\ref{climate}). While
$R_{in}$ is independent on $T$, both $\alpha$ and $E_I$ should be dependent on the Earth's climate which, in
our simple case, is described by the averaged temperature $T$. In a first approximation, we can think that
$E_I$ is linearly dependent on $T$ (a more complex behaviour does not substantially change the results). The so called
albedo $\alpha$ is also a $T$ dependent quantity. More specifically, we expect $\alpha$ to be quite large
for ice regions. An appropriate way to think of the function $\alpha(T)$ is shown in figure (\ref{fig1}). 
Following the cartoon shown in the same figure, it is clear that there exist more than one stationary solution of 
equation (\ref{climate}), corresponding to three different possible Earth climates. The coldest of the possible
climates is referred to as ``ice cover earth'' climate, i.e. the case where almost all incoming radiation is
reflected back in the space. A careful estimate of the albedo effect and the infrared radiation, obtained by current
observations, shows that the warmest possible climate is close to the current Earth climate (we discard the intermediate
climate simply because it is dynamically unstable). Thus equation (\ref{climate}) gives us a quite amazing result as
it explains our present climate state. Note that an increase of $CO_2$ in the atmosphere is equivalent, in our simple model,
to a change in $E_I$ represented by the thick line in the lower panel of figure (\ref{fig1}). 

Our simple model tells us that there is an interesting feature in climate dynamics represented by the instability of the
intermediate climate state. What is the physical origin of this instability? The answer is related to the complex mechanism
of cloud formation. Clouds by themselves reflect incoming radiation. When the temperature increases, the albedo start to 
increases as
well because more clouds are expected to form. Eventually, for large enough temperature, the amount of cloud formation 
is balanced by precipitation 
leading to a constant value of $\alpha$. For the time being, let us concentrate on the present (stable)
climate. The orbital forcing, discovered by Milankovitch, is a small amplitude modulation of $R_{in}$ in time, i.e.
\be
\label{forcing}
R_{in}(t) = R_{in}^S + A cos(\omega t)
\ee
Let us consider the effect of the Milankovitch forcing in equation (\ref{climate}). If $T$ is always quite close
to the stationary (present) climate $T_0$, we can estimate $\delta T \equiv T-T_0$ as:
\be
\label{effect}
\frac{d \delta T}{dt} = - \frac{1}{\tau} \delta T + A cos(\omega t)
\ee
where $1/\tau$ is the relaxation time of the present climate. More precisely $\tau$ can be estimated by the equation:
$$
\frac{1}{\tau} = -\frac{1}{C_E} \frac{[E_I-(1-\alpha(T))R_{in}]}{dT}|_{T=T_0}
$$
After an appropriate estimate of $A$,
one can easily show using (\ref{effect})
that the effect of the orbital forcing on the present climate is of order $A \tau \sim 0.5 K$, much
smaller than the $10 K$ observed in paleoclimatic records. Thus, it seems that orbital forcing can hardly explain
the glacial/interglacial transition experienced by the Earth climate. 

\begin{figure}
\centering
\epsfig{width=.40\textwidth,file=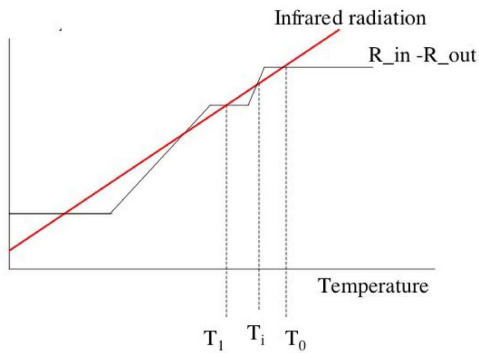}
\caption{The albedo feedback for an energy balance model with three climate states near $T_0$. The model is 
conceptually similar to the one described in figure (\ref{fig1}) with a change in the albedo. }
\label{fig2}
\end{figure}

In order to make progress we need to supply our simple model by new features. An important feature is to assume
that there exist more than one stable climate state close to our present value $T_0$. More precisely, we shall assume
that the albedo feedback mechanim (described in terms of clouds and ice) can eventually lead to a more complex behavior
near our present climate leading to two stable climate states $T_1$ and $T_0$ separated by about $10 K$ see figure (\ref{fig2}). 
This is a rather
ad hoc assumption which is, at any rate, consistent with observations on albedo. Still, this feature does not
really change our conclusion since the orbital forcing is too small to provide a suitable mechanism to 
produce transitions between $T_0$ and $T_1$.

The situation can improve dramatically if we consider the 
following three non trivial statements:
\begin{itemize}
\item{Many complex systems can be described by means of slow variables and fast variables, even if there is no explicit
time scale separation between the two kind of variables. Fast variables may be considered as ``noise'' acting on the
dynamics of slow variables.}
\item{For Earth climate, the averaged temperature in (\ref{climate}) should be considered as a slow variable with
respect to the fast variables due to weather variability.}
\item{When there exist multiple equilibriums, the noise can introduce a long (random) time scale to switch from one equilibrium
to another.}
\end{itemize}

The three statements are no trivial and should be explained carefully. For many decades, the splitting among slow and fast 
variables have been considered a well known feature of molecular motion. At the atomic scale, particles continuously
experience short time interactions with other atoms while the hydrodynamic behavior (the slow variables) is described
by a suitable space average over the molecular chaos. Looking at the fluctuating properties of a turbulent flow, there
exists a well defined scale separation (both in time and in space) from molecular motion and hydrodynamic flow. Nevertheless,
even in a turbulent flow, i.e. disregarding any molecular motion, we can still distinguish large and small scale fluctuations.
In this case, however, there is no spectral gap identifying any scale separation among scales. Therefore, one may be tempted
to state that, in this case, the concepts of fast and slow variables are poorly defined.
It turns out that such a statement
is too limited. There exist experimental systems where the large scale motion can be thought as the slow 
variable superimposed to a noise (turbulent) small scale background acting on the system. Recently \cite{BENZI}, it has been
shown that the above picture can explain qualitatively and quantitatively the numerical results observed in a simplified
model of turbulence. Thus, in most complex systems the separation of large scale, slow variables and
small scales, fast variables does correctly picture the dynamics in a self consistent way.

If we apply the above discussion on the Earth climate (second item), then on the time scale corresponding to the orbital 
forcing,
the day by day weather fluctuations should be considered as noise (fast) variables acting on the climate system. 
This is not just a way of thinking. Disregarding the effect of noise is a major limitation for a correct description
of the physical properties of climate. It took a long time, and it is still on going, for the scientific community to
accept that noise is not a measure of our ``ignorance'' but just a physical feature of most complex systems.

The third item is more delicate and needs to be explained in details. Let us assume that our climate system (although
the discussion is true for any system) shows two stable equilibriums whose temperature difference is $2\Delta T$.
In this case, it is well known that the effect
of the noise can induce transitions between the equilibria. The characteristic time 
$\tau_L$ for the transition can be estimated
as:
\be
\label{time} 
\tau_L  \sim \tau exp(\Delta T^2/\sigma \tau)
\ee
where as in equation (\ref{effect}) $\tau$ is defined as the (fast) relaxation time over one of the stable equilibriium and
$\sigma$ is the variance of the noise (weather fluctuations) acting on the system. Thus our year by year temperature
is assumed to be decomposed in the climatic component $T$ and a fast noise component. In equation (\ref{time}), $\tau_L$ should
be considered an average time to switch from one equilibrium to another. In other words, the
transition time between the two equilibriums $\tau_{eq}$ is a random variable with average value $\tau_L$. Actually,
under some quite generic conditions, we can predict the probability distribution of $\tau_{eq}$ which is
\be
\label{prob_t}
P(\tau_{eq}) = \frac{exp(-\tau_{eq}/\tau_L)}{\tau_L}
\ee
Let me summarize the new feature we have introduced in the simple energy balance model (\ref{climate}). We have
assumed a new form of the albedo which is providing three different climate states around $T_0$, let us call them
$T_1$ (stable), $T_i$ (unstable) and $T_0$. The shape of the new albedo is shown in figure (\ref{fig2}). We assume that
$T_0-T_1 = 10K$ and $T_0-T_i = T_i-T_1$. 
Next we have introduced an external noise which takes into account the dynamics of the fast variables,
whatever they are, not represented by the slow component $T$. We can rescale $T$ near $T_i$ and simplify the equations
by using $T = T_i + \Delta T X$ with $\Delta T = 5^o K$:
\be
\label{climatesimple}
\frac{dX}{dt} = X-X^3 + \sqrt{\sigma} \eta(t)
\ee
where $\eta(t)$ is a gaussian random variable with unit variance ($\langle \eta(t)\eta(t')\rangle = \delta(t-t')$) and $\sigma$
is the noise strength. Notice that we have also rescaled the time unit.
In order to do some computation what we still need is the strength of the noise. We can obtain some information by
the following trick. In our present climate (i.e. $T_0$) we have observational data which tells us about average temperature
fluctuation. Also, we know from General Circulation Models, which is the time correlation for temperature fluctuations.
Thus, we can estimate the strength of the noise by linearizing our climate model around $T_0$
and using the fluctuation-dissipation theorem, i.e. {\it temperature variance = noise strength / (2 relaxation time)}. Once we
know the noise strength $\sigma$ we can compute the average switching time from $T_0$ to $T_1$. It turns out that, by using 
(\ref{time}) the average transition time between the two stable states of our climate model is close to $50,000$ years!
This is a rather surprising and interesting result because it tells us
that there exists a rather long time scale ($50,000$ years) introduced  by the nonlinearity in the
model {\it and} the noise, to be compared against
the quite short time scale of the model, namely the deterministic relaxation time towards one of the two stable
equilibriums which is of the order of $10$ years. However, this is not enough because the characteristic time of 
$50,000$ years is a random time with an exponential probability distribution. Therefore nonlinearity and noise are
not enough to get a periodic behavior of the temperature. 

In order to get a periodic behavior, the idea is to introduce the Milankovitch effect in the system. After a simple
computation one gets the following results:
\be
\label{rs1d}
\frac{dX}{dt} = X-X^3 + A sin(\omega t) + \sqrt{\sigma} \eta(t)
\ee
An appropriate rescaling in terms of physical quantities gives us the value of $A \sim 0.11$ for the Milankovitch terms and
$\omega = 2\pi/10^5 \times 1/10$. Without the noise, the value of $X$ changes periodically in time with an amplitude order $A$.
Going back to the temperature, this implies that, without noise, we have a periodic behavior of temperature with
amplitude $A \Delta T \sim 0.5K$. The situation changes completely when we consider the effect of the noise. 
A correct understanding of the noise effect can be achieved by considering the simplified equation:
\be
\label{example}
\frac{dX}{dt} = X-X^3 + A \sqrt{\sigma} \eta(t)
\ee
i.e. without any time dependency on the external forcing. It is crucial to remind that for any stochastic differential
equation of the form:
\be
\label{stochastic}
\frac{dX}{dt} = -\frac{\partial V}{\partial X} + \sqrt{\sigma} \eta(t)
\ee
the ``equilibrium'' probability distribution $P(X)$ of $X$ is given by:
\be
\label{pdf} 
P(X) = Z_N exp(-2V(X)/\sigma)
\ee
where $Z_N$ is a normalization factor.
In our case $V(X) = -1/2X^2+1/4X^4-AX$ (\ref{example}) and we obtain from (\ref{pdf}) 
\be
P(X;A) = P(X;A=0) exp(2AX/\sigma)
\ee
The probability distribution when $A\ne 0$ changes dramatically the probability distribution when $A=0$. Even for
small $A$ the exponential factor  $ exp(2A/\sigma)$ can be rather large, i.e. order $30$ in our case. The consequence of 
this simple calculation is that the probability to be near $X\sim 1$ is $30$ times larger than to be in $X=-1$. Now it is
quite clear that if $A$ is slowly varying as $cos(\omega t)$ in time than the probability distribution is peaked {\it periodically}
in $\pm 1$ with almost the same period of $2\pi/\omega$. This is the very essence of stochastic resonance.

\begin{figure}
\centering
\epsfig{width=.50\textwidth,file=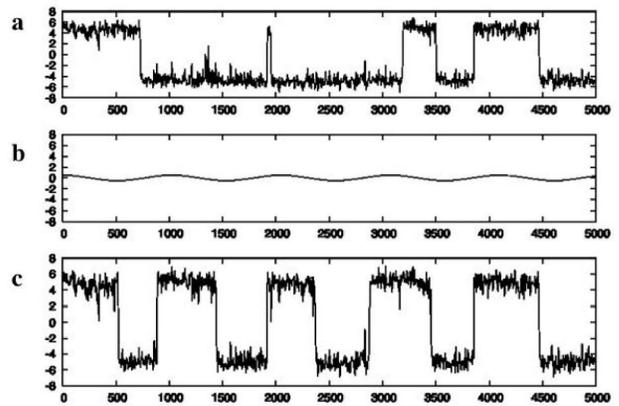}
\caption{Numerical simulation of equation (\ref{rs1d}) in three different cases. In upper panel, we show the solution
with the noise and $A=0$. In the middle panel we show the perturbing forcing $Acos(2\pi \omega t)$ while in the lower 
panel we show the solution with $A\ne 0$. The effect of the small periodic forcing is to synchronized the random switching
from one climate states to the other, i.e. to get a stochastic resonance.}
\label{fig3}
\end{figure}

\bigskip

In figure (\ref{fig3}) we show the numerical simulation of (\ref{rs1d}) and in particular the upper panel is $X(t)$ for 
$A=0$, the pnnel in the center shows the behaviour of $A(t)$ while the lower panel is the solution of (\ref{rs1d}). As we can see,
the effect of the {\it small} periodic perturbation is to synchronize the ``random'' behavior of $X$ and to produce an almost
periodic output. In some sense, stochastic resonance is a counter intuitive phenomenon
because without noise the system shows a small
amplitude modulation around one of the two stable steady states $\pm 1$, while adding the noise we obtain a large amplitude
effect with the same period. 

\bigskip

\begin{figure}
\centering
\epsfig{width=.40\textwidth,file=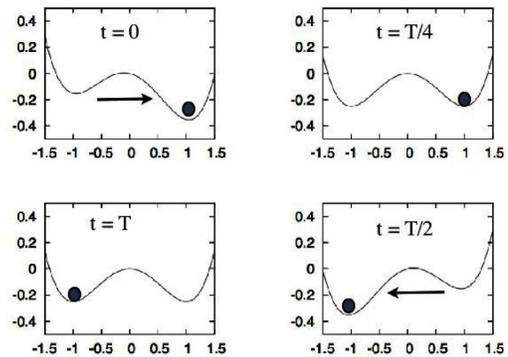}
\caption{A cartoon to explain the mechanism of stochastic resonance. The period of the oscillation is $T$. At time
$t=0$ the climate states has low probability to jump from $+1$ to $-1$. At time $t=T/2$ the situation is reversed:
the probability no to jump is extremely small. At time $t=T$ we start a new cycle. }
\label{fig4}
\end{figure}

One has to be a little bit carefully in understanding the possible results of (\ref{rs1d}). As we said, the modulation of $A$ in 
time should be slow enough in order to get a stochastic resonance. We can be more quantitative using the following simple argument.
Let us go back to equation (\ref{example}). We can compute the probability distribution of the {\it random} time to go
from $+1$ to $-1$ and we already know that the probability distribution is exponential. Let $\tau_1(+1)$ the average exit
time, then the variance of the random switching time is also $\tau_1(+1)$. Finally we know how to relate $\tau_1(+1)$ to $A$, namely by
using the expression:
\be
\tau_1(+1) \sim exp(\frac{2}{\sigma}(\Delta V_0 +A))
\ee
where in our case $\Delta V_0 \equiv 1/4$. Thus the requirement we need is that $\tau_1(+1) \ll 2\pi/\omega$ and
$\tau_1(-1) \gg 2\pi/\omega$ where $\tau_1(-1)$ is the average exit time {\it starting} with the initial condition 
$X=-1$. The two requirements implies 
\bea
\label{c1}
exp(2\frac{\Delta V_0}{\sigma}) \sim \frac{2\pi}{\omega} \\
\label{c2}
exp(\frac{2}{\sigma}(\Delta V_0 -A)) \ll \frac{2\pi}{\omega} 
\eea
The meaning of the above inequalities is represented in the cartoon of figure (\ref{fig4}). 
The time is in the counterclockwise starting with the left top panel.
We start at $X=1$ and as time goes by, the system at $t=T/2$ has a very high probability to jump in $X=-1$ if (\ref{c2}) is satisfied.
Then at $t=T$ we have the same effect for the transition $X=-1 \rightarrow X=1$. The two conditions (\ref{c1},\ref{c2}) tell us
that if the noise is too small then we will never see anything and, on the other hand, if the noise is too large transitions
will happen regardless of the external periodic forcing $A$. Thus, there is a range of $\sigma$ (i.e. the resonant range) where
we obtain periodic transitions {\it in phase} with the external forcing. 

\bigskip

\begin{figure}
\centering
\epsfig{width=.40\textwidth,file=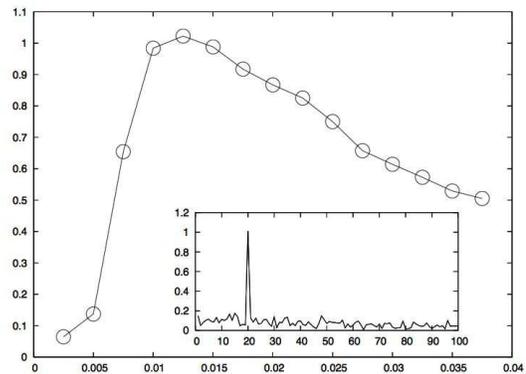}
\caption{The figure shows the Fourier amplitude $|F_X(\nu=\omega)|^2$ for the solution of equation (\ref{rs1d}) for
different values of the noise amplitude $\sigma$. In the insert we show $|F_X(\nu)|^2$ for $\sigma=\sigma_R$, i.e.the 
optimal noise for which the Fourier amplitude at $\nu=\omega$ is maximum. }
\label{fig5}
\end{figure}

One can highlight the resonance effect also by using another kind of variable. Let $X(t)$ the solution of (\ref{rs1d}) and
let us denote $ F_X(\nu) $ its Fourier transform. Then $|F_X(\nu )|^2 $ represents the power spectrum of $X$. We should expect
that $|F_X(\nu=\omega)|^2$ is a function of the noise with some maximum value for the ``resonant noise'' $\sigma_R$. 
This is indeed the case as shown in figure (\ref{fig5}), where 
$|F_X(\nu=\omega)|^2$ is plotted against the $\sigma$, while in the insert we show $|F_X(\nu)|^2$ for $\sigma=\sigma_R$. 

\bigskip

Let me know go back to the climate. As we have seen, the effect of a small periodic forcing in our ``simplified'' climate model is
to produce a periodic glacial to interglacial transition in phase with the Milankovitch forcing. This is a quite good new. However,
the signal shown in figure (\ref{fig3}) poorly compares against the observed behavior of the ``proxy'' earth temperature. 
The latter 
shows a kind of saw tooth behavior with strong asymmetries between cold and warm period. Thus it seems that our theory
cannot represent the real climate. Such a negative statement, however, is somehow misleading simply because
the model so far used, namely the energy balance model (\ref{climatesimple}), includes only the simplest possible feedback due
to radiation. More sophisticated effects, like the temperature-precipitation effect, can change the situation. 
Based on the above discussion we can make the following statement:
\begin{itemize}
\item{stochastic resonance is able to show that the external ``weak'' forcing due to Milankovitch cycle can be amplified by
internal non linear dynamics of the climate;}
\item{climate dynamics shows a long time scale behavior due to non linear feedback;}
\item{noise can develop long time scale behavior if there exists multiple equilibriums;}
\end{itemize}
The most important feature coming out from our analysis is that it is crucial to understand the non linear interaction between
fast a and slow time variables in climate models. Time scales of order $1000$ years or more, do no rule out the importance
of the dynamics of fast variables which usually characterize atmospheric and/or oceanic circulations'. This, I think, is an
important feature even in more sophisticated and/or realistic climate models.

\bigskip

\begin{figure}
\centering
\epsfig{width=.40\textwidth,file=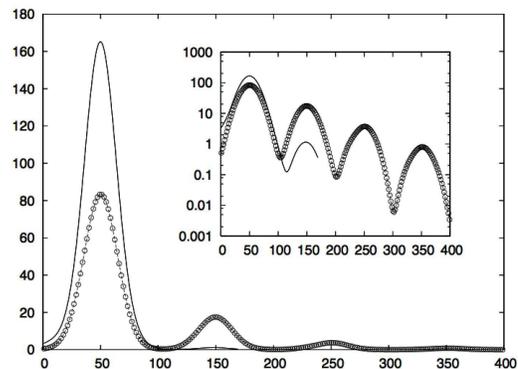}
\caption{Probability distribution of the random switching time $\tau$ for two different noise amplitude obtained
as the solution of equation (\ref{rs1d}). The amplitude and the period of the forcing is the same. Line refers 
to $\sigma_R$ (see figure (\ref{fig5}) while symbols refer to $\sigma = 0.8 \sigma_R$. In the insert we show
the same figure in log-lin plot. It is evident that for $\sigma$ smaller than $\sigma_R$ the probability
distribution of $\tau$ is ``quantized'' at integer values of $(2n+1)T/2$. }
\label{fig5b}
\end{figure}

The mechanism of stochastic resonance has been recently invoked to explain the observed climatic variation on a shorter time scale
with respect to the Milankovitch forcing. During the last $100,000$ years, the earth climate exhibits abrupt changes from relatively
cold to warm states, different to those observed in the $100,000$ year cycles. According to Alley et.al. \cite{ALLEY}, 
the climate changes 
occurred on time intervals which are multiple of $1,500$ years, i.e. the time to switch from one climate state to the other
is either $1,500$ or $3,000$ or $4,500$ and so on. This rather peculiar behavior calls for an explanation which, not surprisingly,
can be given in terms of stochastic resonance. As we previously discussed for equation (\ref{rs1d}), the effect of the periodic
forcing is to decrease and increase (depending on the time $t$) the probability of a switching from a climate state to the other.
With reference to figure (\ref{fig4}), if at time $t=T/2$ the system does not switch, then it takes a full period $T$ for the 
possibility to make a jump. This is certainly the case if the noise intensity is smaller than $\sigma_R$. It follows that 
the probability of the exit time is not peaked around $T/2$ but it is {\it quantized} with maxima in $(2n+1)T/2$, $n=1,2,...$. 
To show that our analysis is correct, in figure (\ref{fig5b}) 
we plot the probability distribution of the time  to switch  for $\sigma = 0.8 \sigma_R$ (symbols)
and for $\sigma = \sigma_R$ (continuous lines). The quantization effect is quite clear as predicted by our simple analysis.

\bigskip
It appears that the observation of Alley et.al. may be explained by the mechanism of stochastic resonance as it was pointed
out by the same authors. Actually, a theoretical analysis of a simplified ocean/atmospheric model, due to Ganopolski and
Rahmstorft \cite{PRLGR} shows that the effect of a periodic forcing in freshwater input over the North Atlantic Ocean combined with
a suitable stochastic forcing, produces a stochastic resonance in cold/warm climate changes as observed in real data. In
the  Ganopolski and Rahmstorft model, 
the two
climate states, between which one has climate transitions, 
 represent two different thermoaline circulation in the Norht Atlantic Ocean, which
are responsable for sea ice growth and destroy. 

It is unclear, at this stage, whether stochastic resonance is or is not the possible explanation of the Alley et.al. data and
more research is needed to be done for a reasonable assessment. At any rate, it is important to remark that abrupt climate changes
as those observed during the Milankovitch cycle or over the last $100,000$ years are hard to be explained without the idea
of transition between multiple equilibriums triggered by the noise, which is 
{\it the} basic feature of stochastic resonance.

\section{3. Stochastic Resonance in complex systems}

In the last ten to fifteen years the mechanism of stochastic resonance has been widely applied to a number of different
physical and biological systems. Among them, the most striking applications concern neural systems pioneered by Moss and
collaborator \cite{MOSS}. In most cases, experimental results show a remarkable agreement with respect to the
qualitative and quantitative picture represented in figure (\ref{fig5b}). Also, in many applications one does not know the
``equation of motions'' for the system and, actually, periodic perturbation leading to a stochastic resonance are used
as a ``measure" of non trivial cooperative phenomenon in the system under study (this is certainly the case for most
neurophysiological systems). 

\bigskip

\begin{figure}
\centering
\epsfig{width=.40\textwidth,file=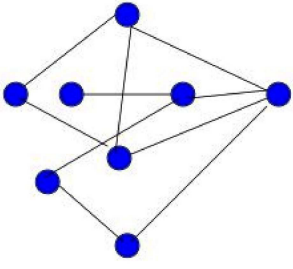}
\caption{A cartoon of our ``complex'' system. Each node satisfies equation (\ref{network}). Lines describes
the connectivity matrix $L_{ij}$.}
\label{fig10}
\end{figure}

When dealing with complex systems, one can find non trivial behavior of the stochastic resonance. In this section I show
that this is the case for a network system composed by some ``node' $\psi_i$, $i=1,...N$, see figure (\ref{fig10}).

Just to simplify the discussion,
I assume that the dynamics of each node is controlled by the same equation and that the connectivity of the system is described
by the matrix $L_{ij}$:
\be
\label{network}
\frac{d\psi_i}{dt} = m \psi_i -g\psi_i^3 + L_{ij} \psi_j + \sqrt{\sigma} \eta_i
\ee
where $m > 0$ and $g$ are real constant.
Also, I will assume that $\sum_i L_{ij} \psi_j \equiv 0$ which clearly can be done without lack of generality, and
finally, that $\sum_i L_{i,j} \psi_i \psi_j >0$, i.e. the connectivity matrix does not introduce any ``instability'' in the
dynamics of the system.
The noise $\eta_i$ is $\delta$-correlated in time and $\langle \eta_i \eta_j \rangle = \delta_{ij}$. 

\bigskip

The complexity of our problem, so to speak, is introduced by the matrix $L_{ij}$. We want now to study the behavior of
(\ref{network}) when an external periodic forcing is added to the system, i.e. 

\be
\label{ris2dA}
\frac{d\psi_i}{dt} = m \psi_i -g\psi_i^3 + L_{ij} \psi_j + \sqrt{\sigma} \eta_i + A_i cos(2\omega t)
\ee
In order to simplify our work, let us focus on the ``average'' variable $\Phi = N^{-1} \sum_i \psi_i$. One should expect
that for long enough period $T=2\pi/\omega$ and a suitable noise $\sigma_R(m,g)$ a stochastic resonance can be observed. Note that
I have defined $\sigma_R$ as an explicit function of the variables $m$ and $g$. By averaging (\ref{network}) we obtain
\be
\frac{d\Phi}{dt} = m \Phi - g \langle \psi_i^3 \rangle + \sqrt{\epsilon} \eta 
\ee
where $\sqrt{\epsilon} \equiv N^{-1}\sum_i \sqrt{\sigma}$ and $\langle ... \rangle \equiv N^{-1} \sum_i ....$. I assume that
$\epsilon$ is independent of $N$, i.e. $\sigma$ is chosen in such a way that $\epsilon$ is a fixed quantity.
The difficulty 
is to compute the term $\langle \psi_i^3 \rangle$. For this purpose, we define $\phi_i$ such that $\psi_i = \Phi + \phi_i$, i.e.
$\phi_i$ are ``deviation'' of $\psi_i$ from $\Phi$. Then we have 
$\langle \psi_i^3 \rangle = \Phi^3 + 3 g \Phi \langle \phi_i^2 \rangle $. This expression is correct as far as we
can neglect the term $\langle \phi_i^3 \rangle$ which, in most cases, is a good first approximation. Putting all together,
we obtain:
\be
\label{LG}
\frac{d\Phi}{dt} = (m-  3 g \langle \phi_i^2 \rangle)\Phi - g \Phi^3  + \sqrt{\epsilon} \eta 
\ee
As we can see, the effect of complexity, i.e. the connectivity matrix $L_{ij}$, introduces a change in the linear term which
now become a {\it time dependent function }. It can happen that $ (m- 3 g \langle \phi_i^2 \rangle) \le 0$ and if this is the case,
transition between the two states (whatever they are) occur with a mechanism completely different with respect to what we
discussed in the previous section. Is this the case? Everything depends on the quantity $\langle \phi_i^2 \rangle$ and
therefore on the connectivity matrix $L_{ij}$.

\begin{figure}
\centering
\epsfig{width=.50\textwidth,file=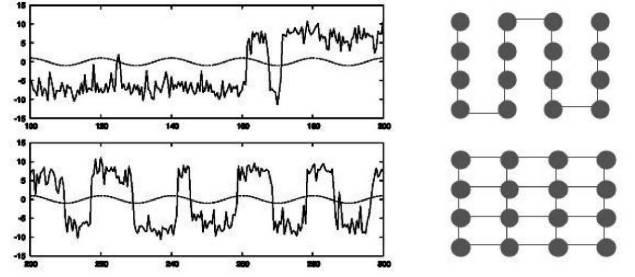}
\caption{Numerical simulation of equation (\ref{network}) with the same parameters $m,g,\sigma$ and
two different topology (shown in the figure): one dimensional lattice (upper panel), two dimensional lattice
(lower panel). The amplitude and the period of the periodic forcing is the same for both cases. While the two dimensional
lattice shows stochastic resonance, the same is not true for the one dimensional case. }
\label{fig11}
\end{figure}

In order to make progress, let me first discuss what are the ``statistical stable''  equilibriums  
of (\ref{LG}). Let us define
\be
\Phi_0^2 \equiv \frac{m-3g\langle \phi_i^2 \rangle}{g}
\ee
We should expect that the statistical stable equilibriums are $\pm \Phi_0$. Then fluctuations around the equilibrium, which
in first approximation we can describe as $\phi_i$, satisfy the equation:
\be
\frac{d\phi_i}{dt} = - \alpha \phi_i + L_{ij} \phi_j + \sqrt{\sigma} \eta_i
\label{fluct}
\ee
where 
\be
\label{alpha}
\alpha \equiv  (m-3g\langle \phi_i^2 \rangle) - 3g\Phi_0^2 = -2 (m-3g\langle \phi_i^2 \rangle)
\ee
Using (\ref{fluct}) we can now estimate $\langle \phi_i^2 \rangle$. Let us define $-\lambda_n$ the eigenvalues
of $L_{ij}$. Then, by using a well known result in the theory of stochastic differential equation, we obtain:
\be
\langle \phi_i^2 \rangle = \sum_i \phi_i^2 = \frac{1}{2}\sum_n \frac{\sigma}{\alpha +\lambda_n}
\ee
In the limit of large $N$, we can define the density of states $\rho(\lambda)$ with the approximation:
\be
\sum_n \rightarrow \int d\lambda \rho(\lambda)
\ee
Putting everything together, we finally have:
\be
\label{phi_2}
\langle \phi_i^2 \rangle = \frac{1}{2}\int d\lambda \frac{\rho(\lambda)}{\alpha +\lambda_n} = 
\frac{1}{2}\int d\lambda \frac{\rho(\lambda)}{2 (m-3g\langle \phi_i^2 \rangle)+\lambda}
\ee
Equation (\ref{phi_2}) is a non linear equation relating the value of $\langle \phi_i^2 \rangle$ to the density of states
$\rho(\lambda)$,i.e. to the connectivity matrix $L_{ij}$. Once we have $\langle \phi_i^2 \rangle$ we can comput $\Phi_0$. 
The tricky part of our problem is that we have computed
$\langle \phi_i^2 \rangle$ when $\langle A \rangle =0 $, i.e. with no
external forcing. If we now have an external forcing, all our computation for $\Phi_0$ and $\alpha$ should change taking into
account $A$. One can compute perturbatively the effect in power of $A$ and take the first order for small $A$. The computation
are done for a special case in \cite{BENZISUTERA}. The final results is that everything goes as in the theory 
discussed in section 2 but with a renormalized value of $A$,i.e.
\be
\label{AR}
A_R \equiv A(1+\frac{3gD}{1-2Dg})
\ee
where 
\be
\label{D}
D \equiv \sigma \int d\lambda \frac{\rho(\lambda)}{(\alpha_0 + \lambda)^2}
\ee
and $\alpha_0$ corresponds (\ref{alpha}) for $A=0$. For any practical purpose our result means the following. Let us imagine
a network described by equation (\ref{network}). Then the effect of a periodic forcing with amplitude $A$ on the 
average $\Phi$ is equivalent to a one dimensional problem (similar to (\ref{climatesimple}) with a renormalized
amplitude (\ref{AR}). The connectivity matrix fixes the value of the renormalization by (\ref{D}). Thus, depending
on $L_{ij}$, the effect of stochastic resonance can be enhanced or depressed.  In figure (\ref{fig11}) we demonstrate our
result in a simple case. We consider the same system (i.e. same value of $m,g,A$ and $\sigma$) for connectivity
matrix topologically equivalent to a one dimensional lattice
(upper panel), and a bidimensional lattice. In the latter case the system shows stochastic resonance while the
same is not true for the former case. 

\bigskip

One main conclusion that we can outline from our discussion is particular relevant for climate theory. The
physical meaning effect of an external 
forcing can be drastically different depending on the feedback in the system (in our
case the matrix $L_{ij}$). There are cases, as we have seen, for which a small forcing or a small noise 
can trigger transitions and a naive computation of the relevant quantities (i.e. neglecting the effect discussed
in this section) can lead us to wrong results.
I want to argue that the above conclusions is relevant
for other physical and biological systems.

\section{4. Brownian Motors and Stochastic Resonance}

Up to now we have discussed the case of stochastic resonance in a more or less traditional fashion, i.e. when multiple
equilibriums exist. We want now to understand whether there are other physical problems where a stochastic-like resonance 
can be of interest. One possible candidate, that I discuss in this section, is the case of Brownian motors.

\begin{figure}
\centering
\epsfig{width=.40\textwidth,file=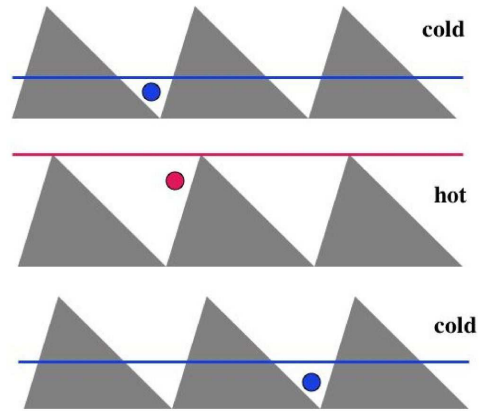}
\caption{A simple model of Brownian motor.Going from the top to the bottom of
when the temperature is low (cold) a particle moving in the ratchet potential is
trapped in one of the minima. Rising the temperature, the particle becomes almost free to move and, because its initial
condition (minimum) is close the maximum on its right, the particle has more probability to go towards the right minimum. 
If the temperature is now decreasing the particle has more probability to be trapped in the right minimum with respect to
be trapped in the minimum on the left  }
\label{b1}
\end{figure}
The very idea of Brownian motors goes back to Feynman \cite{FEYNMAN} and it can be summarized by 
saying that we want to use the energy of the thermal noise in order
to make work. This can be done, consistently with the law of thermodynamics, if one considers a ratchet potential with
a periodic time behavior of the temperature, see figure (\ref{b1}). Going from the top to the bottom of
figure (\ref{b1}), we see that when the temperature is low (cold) a particle moving in the ratchet potential is
trapped in one of the minima. Rising the temperature, the particle becomes almost free to move and, because its initial
condition (minimum) is close the maximum on its right, the particle has more probability to go towards the minimum on the right. 
If the temperature is now decreasing the particle has more probability to be trapped in the right minimum with respect to
be trapped in the minimum on the left. The overall picture is that of a diffusion process with more probability to move
on the right than on the left. Thus, under the combined action of periodic temperature variation and the ratchet potential, 
we can have a non zero drift average velocity of the particle on the right direction: we have been able to extract work
from the temperature field.

\begin{figure}
\centering
\epsfig{width=.40\textwidth,file=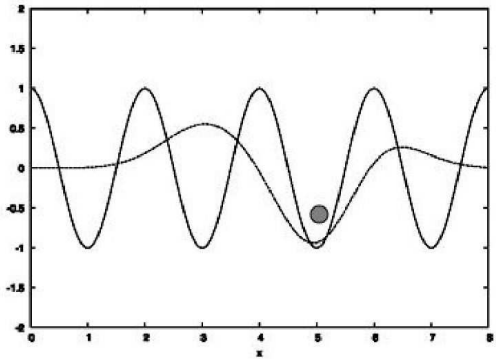}
\caption{The Brownian motor proposed in section 4. The particle moves in a periodic potential and noise is acting on it.
A perturbing potential $\Phi(x-vt)$ localized in space is traveling in the system. }
\label{b3}
\end{figure}
\begin{figure}
\centering
\epsfig{width=.40\textwidth,file=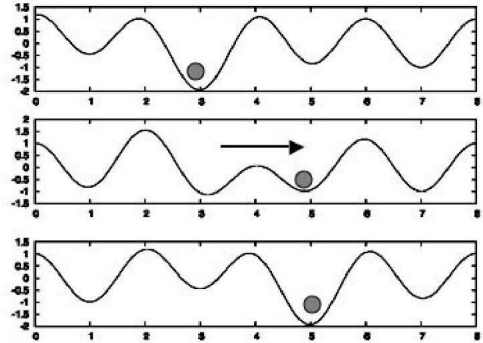}
\caption{The effect of the perturbing potential $ \Phi (x-vt)$. From the top to the bottom, the particle can
flip only in the positive x-direction as the potential travels. The flip occurs only for a suitable combination of $v$
and the noise amplitude $\sigma$.}
\label{b2}
\end{figure}

There are many kind of Brownian motors \cite{MOTORS}, 
the one shown in figure (\ref{b1}) is just an example. I will propose here a new kind of 
motors using the very same idea of stochastic resonance to produce a non zero current using the noise. Instead of a ratchet
potential, let me consider a particle which feel a space  periodic potential of period $L$. In the limit of overdamped friction,
the equation of motion of the particle position $x$ is:
\be
\frac{dx}{dt} = -\frac{\partial V}{\partial x} + \sqrt{\sigma} \eta(t) \ \ \  V(x) = V_0 cos(2\pi x/L)
\ee
We now introduce a time dependent forcing $F(x-vt)$ in the form of a moving localized perturbation as shown in figure (\ref{b3}).
Actually we can thing of $F(\xi)$ as the effect of a moving potential perturbation, i.e.
\be
F(x-vt) = - \frac{ \partial \Phi(x-vt)}{\partial x}
\ee
Without the effect of $F$, the particle performs a diffusive process among the minima of $V$. The characteristic time
$\tau_L$ to jump on the right or on the left can be computed by using the same method of section 2. The action of the
localized potential can introduce a new feature which is illustrated in figure (\ref{b2}).

\begin{figure}
\centering
\epsfig{width=.40\textwidth,file=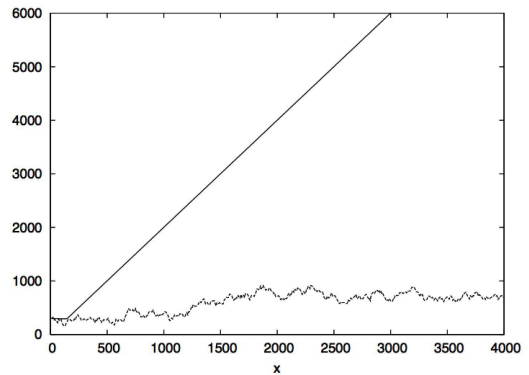}
\caption{The figure shows $x(t)$ with (straight line) and without (dotted line) $\Phi(x-vt)$ 
for a properly chosen value of $\sigma$. The
initial particle position is $300L$ and the initial perturbing potential is localized in $x=0$. }
\label{b4}
\end{figure}

If the speed $v$ of the perturbing potential $\Phi$ is properly chosen, then the particle has the possibility to jump
on the right but not on the left when the perturbing potential $\Phi$ is in a given position. Then the particle remains
trapped in the new minimum up to the time when the potential is now shifted to the next minimum and so on. In other words,
the particle remains trapped in the moving potential $\Phi$ and it moves ``balistically'' at the same speed $v$ of the perturbing
potential.  In figure (\ref{b4}) I show $x(t)$ with and without $\Phi$ for a properly chosen value of $\sigma$. The
initial particle position is $300L$ and the initial perturbing potential is localized in $x=0$.

When $\Phi$ is close to $x=300$, the particle remains trapped and it moves with the speed $v$ for a rather long distance.
According to our knowledge of stochastic differential equation, at each step the particle is following the potential
with a probability close to $1$ but not exactly equal to $1$. It follows that the particle is trapped in the potential 
$\Phi$ for a distance $D_x$ which is a random variable (it is the analogous of the random time $\tau_L$ discussed in 
detail in section 2).

\bigskip

\begin{figure}
\centering
\epsfig{width=.40\textwidth,file=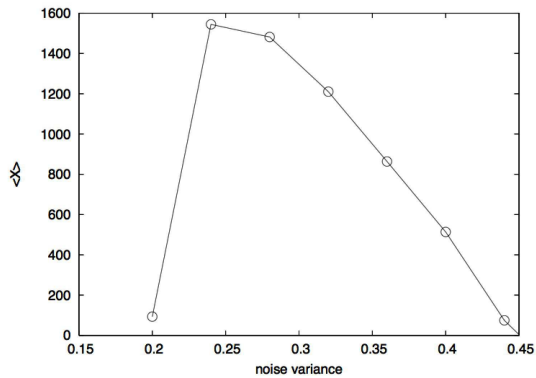}
\caption{The quantity $\la D_x \ra$ as a function of $\sigma$ }
\label{b5}
\end{figure}

\begin{figure}
\centering
\epsfig{width=.40\textwidth,file=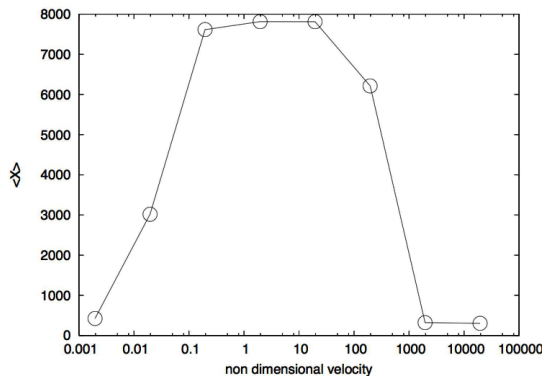}
\caption{The quantity $\la D_x \ra$ as a function of $v$  }
\label{b6}
\end{figure}

As for the stochastic resonance, if the noise is to small or too large, $\Phi$ has no effect on the particle. The same is
true if the velocity is too small or too large. We can check these qualitative statements in the following way. We consider
an ensemble of $N$ particles and we study the average distance $\la D_x \ra$ as a function of noise variance $\sigma$ and the
non dimensional velocity $ v \tau_L / L$. This is done in figures (\ref{b5}) and (\ref{b6}) respectively.

In both cases, the effect of a resonant like phenomena is quite clear. Only for properly chosen values of 
$\sigma $ and $v$ we obtain the maximum of $ \la D_x \ra$. Some recent experimental results \cite{WAVE} seem
to confirm the numerical analysis done in this section.

\section{5. Conclusions}

In the last 30 years the scientific community did learn many different features on complex systems, starting by the
pioneering works of Lorenz, Ruelle, Mandelbrot and others. In many cases, new ideas and tools have been introduced
in order to ``measure'' complexity in an appropriate way. These tools 
can be used to reveal different features of underlying physical or biological mechanisms. In some sense, stochastic
resonance is also a tool because it allows us to understand whether or not non linear effects can act in a cooperative way
with the complex and chaotic behavior of a given system. On the other hand stochastic resonance is a mechanism in the
full meaning of the word because it allows to get large effect from a small amplitude perturbation. There have been
and still there are many applications of stochastic resonance in problems dealing with the
amplification of signal to noise ratio, a quite traditional
engineering problem.

\bigskip

In this paper I have reviewed some known and less known features of stochastic resonance. Some simple conclusions can be 
made.
\begin{itemize}
\item{ Stochastic resonance is counter intuitive phenomenon: it is not trivial that adding noise to a system we can enhance
the deterministic periodic behavior.}
\item{ Stochastic resonance is a robust mechanism observed in many physical and biological systems. The notion of stochastic resonance
is now cross disciplinary and new applications are found every year.}
\item{We learn a lot in applying stochastic resonance in the theory of climatic change. As I mentioned in section 2, it is 
a crucial step to understand that fast variables cannot be simply ignored in the study of long term climatic change, which
is, overall, the basic idea introduced by stochastic resonance.}
\end{itemize}

{ \bf Acknowledgments}
I would like to thank many friends and colleagues for their help in early days of my research in stochastic resonance. 
Among them a special mention must be given to Giorgio Parisi, Alfonso Sutera and Angelo Vulpiani. Also, I am indebted with
Micheal Ghil for his help and contribution. This paper represents the
talk I gave on april 2006 for the L.F. Richardson lecture at EGU annual meeting in Vienna. I would like to thank the
scientific committee of the EGU 
Richardson' medal and in particular D. Schertzer for his help and assistance.
The paper has been written 
during a period of stage at the University of Chicago, Flash Center. I thank L. Kadanoff, D. Lamb, B. Fischer for their
kind invitation and support.

\end{document}